\documentclass[12pt]{article}
\usepackage{graphicx}
\usepackage[varg]{txfonts}
\usepackage{cite} 
\usepackage[scaled=0.9]{helvet}

\usepackage{url}  
\usepackage{array}

\evensidemargin\oddsidemargin
\parskip=\medskipamount
\belowcaptionskip=\medskipamount

\makeatletter
\long\def\@makecaption#1#2{%
  \vskip\abovecaptionskip
  \sbox\@tempboxa{\small{\bfseries #1} \  #2}%
  \ifdim \wd\@tempboxa >\hsize
    \small{\bfseries #1} \  #2\par
  \else
    \global \@minipagefalse
    \hb@xt@\hsize{\hfil\box\@tempboxa\hfil}%
  \fi
  \vskip\belowcaptionskip}
\renewcommand\section{\@startsection {section}{1}{\z@}%
      {-3.25ex\@plus -1ex \@minus -.2ex}%
      {1ex \@plus .2ex}%
      {\normalfont\large\sffamily\bfseries}}
\renewcommand\subsection{\@startsection{subsection}{2}{\z@}%
      {-3ex\@plus -1ex \@minus -.2ex}%
      {0.5ex \@plus .2ex}%
      {\normalfont\normalsize\sffamily\bfseries}}
\renewcommand\subsubsection{\@startsection{subsubsection}{3}{\z@}%
      {-3ex\@plus -1ex \@minus -.2ex}%
      {0.25ex \@plus .2ex}%
      {\normalfont\normalsize\sffamily\bfseries}}
\renewcommand\paragraph{\@startsection{paragraph}{4}{\z@}%
      {3ex \@plus1ex \@minus.2ex}%
      {-1em}%
      {\normalfont\normalsize\sffamily\bfseries}}
\renewcommand\subparagraph{\@startsection{subparagraph}{5}{\z@}%
      {1ex \@plus.5ex \@minus .2ex}%
      {-1em}%
      {\normalfont\normalsize\sffamily\bfseries}}
\makeatother

\def\qsqmax{q^2_\mathrm{max}}
\def\sth{s_\mathrm{th}}

\def\modvub{|V_{ub}|}
\def\n#1e#2n{#1\times10^{#2}}
\def\mi{\mathrm{i}}

\def\gev{\,\mathrm{GeV}}
\def\mev{\,\mathrm{MeV}}
\def\d{\mathrm{d}} 
\def\fom#1{f^{\mbox{\scriptsize Omn\`es}}(q^2_#1,f_0,f_1,f_2,f_3)}

\def\vubresult{\n(4.02\pm0.35)e-3n}
\def\fzeroresult{0.215\pm0.024}
\def\vubfresult{\n(8.7\pm1.0)e-4n}
\def\nolcsr{$\modvub = \n4.24(40)e-3n$, $f^+(0)=0.166(31)$}
\def\scet{$\modvub = \n4.24(40)e-3n$, $f^+(0)=0.167(27)$}
\def\lcsrscet{$\modvub = \n3.96(34)e-3n$, $f^+(0)=0.210(22)$}

\begin{document}
\begin{flushright}
SHEP--0624
\end{flushright}

\begin{center}\Large\bfseries\sffamily
Extracting $\modvub$ from $B\to\pi l\nu$ Decays Using a
Multiply-subtracted Omn\`es Dispersion Relation
\end{center}

\begin{center}
\textbf{\textsf{Jonathan M Flynn${}^\mathrm{a}$ and Juan
  Nieves${}^\mathrm{b}$}}\\[2ex]
${}^\mathrm{a}$School of Physics and Astronomy, University of
  Southampton\\
  Highfield, Southampton SO17~1BJ, UK\\
${}^\mathrm{b}$Departamento de F\'isica At\'omica, Molecular y
  Nuclear, Universidad de Granada,\\
  E--18071 Granada, Spain
\end{center}
\medskip

\begin{quote}
\begin{center} 
\textbf{\textsf{Abstract}}
\end{center}
We use a multiply-subtracted Omn\`es dispersion relation for the form
factor $f^+$ in $B\to\pi$ semileptonic decay, allowing the direct
input of experimental and theoretical information to constrain its
dependence on $q^2$, thereby improving the precision of the extracted
value of $\modvub$. Apart from these inputs we use only unitarity and
analyticity properties. We obtain $\modvub=\vubresult$, improving the
agreement with the value determined from inclusive methods, and
competitive in precision with them.
\end{quote}

\section{Introduction}

The magnitude of the element $V_{ub}$ of the Cabibbo-Kobayashi-Maskawa
(CKM) quark mixing matrix plays a critical role in testing the
consistency of the Standard Model of particle physics and, in
particular, the description of CP violation. Any inconsistency could
be a sign of new physics beyond the standard model.$V_{ub}$ is
currently the least well-known element of the CKM matrix and
improvement in the precision of its determination is highly desirable
and topical.

$\modvub$ can be determined using inclusive or exclusive charmless
semileptonic $B$ decays. The inclusive method has historically
provided a more precise result, but recent
experimental~\cite{Athar:2003yg,Aubert:2005cd,Hokuue:2006nr,Aubert:2006ry} and
theoretical
developments~\cite{Arnesen:2005ez,Becher:2005bg,Hill:2006ub,Gulez:2006dt,Okamoto:2005zg,Mackenzie:2005wu,LCSR_04_BZ}
are allowing the exclusive method to approach the same level of
precision. It is important to check the compatibility or otherwise of
results from the two methods, which currently agree only at the edge
of their respective one-standard-deviation errors.

In principle, a comparison using a calculated form factor, which
contains the nonperturbative QCD input, at a single value of $q^2$
with an experimentally determined differential decay rate at the same
$q^2$ would allow the extraction of $\modvub$. In practice,
experimental results are available for the differential decay rate
integrated over $q^2$
bins~\cite{Athar:2003yg,Aubert:2005cd,Hokuue:2006nr,Aubert:2006ry},
providing shape information, while theoretical calculations of the
form factors provide normalisation at a set of $q^2$ values.

Lattice QCD, originally in the quenched
approximation~\cite{Latt_96,Latt_98,Latt_98_nrqcd,Latt_01_nrqcd,
Latt_00,Latt_01,Latt_01_bis} and more recently using dynamical
simulations~\cite{Gulez:2006dt,Okamoto:2005zg,Mackenzie:2005wu},
provides form factor values for the high $q^2$ region because of the
limitation on the magnitude of spatial momentum components. Light cone
sumrules (LCSR), in contrast, determine the form factors in the low
momentum transfer region at or near
$q^2=0$~\cite{LCSR_98,LCSR_98_2,LCSR_00,LCSR_01_BZ,LCSR_Hqet_01,LCSR_02,
LCSR_Hqet_03,LCSR_03,LCSR_04_BZ}.

To combine the theoretical and experimental information requires a
parameterization of the relevant form factor, $f^+(q^2)$, ideally based
on general principles. A dispersion relation motivates
parameterizations by the $B^*$ pole plus a sum of effective poles
(restricted and/or simplified sums are used
in~\cite{Becirevic:1999kt,LCSR_04_BZ}), with a constraint imposed by
the asymptotic behaviour of $f^+$ at large $q^2$~\cite{Becher:2005bg}.
An alternative parameterization stems from the fact that the $B\pi$
contribution can no more than saturate the production rate of all
states coupling to the $\bar u \gamma^\mu b$ current. The latter
`dispersive bound' was first used in this context to bound the form
factors~\cite{Lellouch:1995yv,Fukunaga:2004zz}. More recently, it has
been used to motivate a particular functional form which makes it easy
to test consistency with the
bound~\cite{Arnesen:2005ez,Becher:2005bg,Hill:2006ub}.

Here, we use a multiply-subtracted Omn\`es dispersion relation to
obtain a parameterization of the form factor based only on the
Mandelstam hypothesis~\cite{Ma58} of maximum analyticity, unitarity
and an application of Watson's theorem~\cite{Wa54}. The latter theorem
implies that $f^+$ has the same phase as the elastic $\pi B\to\pi B$
scattering $T$-matrix in the $J^P=1^-$, isospin-$1/2$ channel,
\begin{eqnarray}
\frac{f^+(s+\mi\epsilon )}{f^+(s-\mi\epsilon )}
 = \frac{T(s+\mi\epsilon )}{T(s-\mi\epsilon )}
 = e^{2\mi\delta(s)},\quad s > \sth \equiv (m_B+m_\pi)^2,
\qquad 
T(s) = \frac{8\pi\mi s}{\lambda^{1/2}(s)}
       \left(e^{2\mi\delta(s)}-1\right).
\label{eq:wat}
\end{eqnarray}
The $(n{+}1)$-subtracted Omn\`es representation for $f^+(q^2)$, with
$q^2<\sth$, reads (for more details see the discussion and example in
the appendix of~\cite{NRCQM-bpi}):
\begin{eqnarray}
  f^+(q^2) &=& \bigg(\prod_{i=0}^n\left[f^+(s_i)\right]^{\alpha_i(q^2)}\bigg)
  \exp\bigg\{I_\delta(q^2;\,s_0,\ldots,s_n)
             \prod_{j=0}^n(q^2-s_j)
      \bigg\}, \label{eq:omnes} \\ 
I_\delta(q^2;\, s_0,\ldots,s_n) &=&
  \frac1{\pi}\int_{\sth}^{+\infty}
  \frac{\d s}{(s-s_0)\cdots(s-s_n)}\,\frac{\delta(s)}{s-q^2},
\label{eq:phase-integral}\\
\alpha_i(s) &\equiv& \prod_{j=0, j\neq i}^n
        \frac{s-s_j}{s_i-s_j},\qquad
\alpha_i(s_j)=\delta_{ij},\qquad
\sum_{i=0}^n \alpha_i(s) = 1.
\end{eqnarray}
This representation requires as input the elastic $\pi B \to \pi B$
phase shift $\delta(s)$ plus the form factor values $\{f^+(s_i)\}$ at
$n+1$ positions $\{s_i\}$ below the $\pi B$ threshold. As the
subtraction points coalesce to some common $s_0$, our result reduces
to an expression involving the form factor and its derivatives at
$s_0$ (such a representation was used successfully to account for
final state interactions in kaon decays~\cite{Omnes_01_kaon}). The
asymptotic behaviour of $f^+$ imposes a constraint on the subtractions
(when more are used than needed for
convergence)~\cite{Bourrely:2002fe}, but we keep in mind that we will
apply the representation above only in the physical region of $q^2$
for $B\to\pi$ decay.

As the number of subtractions increases the integration region
relevant in equation~(\ref{eq:phase-integral}) shrinks. If this number
is large enough, knowledge of the phase shift will be required only
near threshold. Close to threshold, the $p$-wave phase shift behaves
as
\begin{equation}
\delta(s) = n_b \pi - p^3 a + \cdots \label{eq:levin}
\end{equation}
where $n_b$ is the number of bound states in the channel (Levinson's
theorem~\cite{MS70}), $p$ is the $\pi B$ center of mass momentum and
$a$ the corresponding scattering volume. In our case $n_b=1$ if we
consider the $B^*$ as a $\pi B$ bound state. Moreover, $m_{B^*}^2$ is
not far from $\sth$. We will perform a large number of subtractions so
that approximating $\delta(s)\approx \pi$ in
equation~(\ref{eq:phase-integral}) is justified. The factor $I_\delta$
can then be evaluated analytically and we find an explicit formula for
$f^+(q^2)$ when $q^2 < \sth$,
\begin{equation}
  f^+(q^2) \approx \frac1{\sth -q^2} \prod_{i=0}^n
  \left[f^+(q^2_i)(\sth-q^2_i)\right]^{\alpha_i(q^2)}, \qquad n \gg 1.
  \label{eq:omn_th}
\end{equation}
This amounts to finding an interpolating polynomial for
$\ln[(\sth-q^2)f^+(q^2)]$ passing through the points
$\ln[(\sth-q_i^2)f^+(q_i^2)]$ at $q^2_i$.

In equation~(\ref{eq:omnes}) we have assumed that $f^+$ has no poles. In
the Omn\`es picture, the $B^*$ is treated as a bound state and is
incorporated through the phase-shift integral. Since $m_{B^*}^2$ is
close to $\sth$, the $B^*$ pole's influence appears in the factor
$1/(\sth-q^2)$ in equation~(\ref{eq:omn_th}). Going beyond the
approximation $\delta(s)=\pi$, the form factor will be sensitive to
the exact position of the $B^*$ pole, since the effective range
parameters (scattering volume, \ldots) will depend on $m_{B^*}$.

In the following we use the explicit formula in
equation~(\ref{eq:omn_th}) with four subtractions\footnote{For four
subtractions, we have checked that there are negligible changes in our
results if the model in~\cite{Omnes_01} for the phase shift is used in
the integral in equation~(\ref{eq:phase-integral}).}. We have performed
a simultaneous fit to $f^+$ values from unquenched lattice QCD and
LCSR calculations, together with experimental measurements of partial
branching fractions. Our main results are:
\begin{equation}
\modvub=\vubresult, \qquad
\modvub f^+(0) = \vubfresult.
\end{equation}
The $9\%$ error for $\modvub$ is competitive with the $7\%$ error
currently quoted for the determination of $\modvub$ from inclusive
semileptonic $B$ decays. Our fitted form factor is consistent with
dispersive constraints~\cite{Arnesen:2005ez,Becher:2005bg}.

\section{Fit Procedure}

The hadronic part of the $B^0\to \pi^- l^+ \nu_l$ decay matrix element
is parametrized by two form factors as
\begin{equation}
\langle \pi (p_\pi) | V^\mu | B (p_B) \rangle =
\left(p_B+p_\pi-q\frac{m_B^2-m_\pi^2}{q^2} \right)^\mu f^+(q^2) +
q^\mu \frac{m_B^2-m_\pi^2}{q^2} f^0(q^2) 
\end{equation}
where $q^\mu = (p_B-p_\pi)^\mu$ is the four-momentum transfer. The
meson masses are $m_B=5279.4\mev$ and $m_\pi=139.57\mev$ for $B^0$ and
$\pi^-$, respectively. The physical region for the squared
four-momentum transfer is $0 \leq q^2 \leq \qsqmax\equiv
(m_B-m_\pi)^2$. If the lepton mass can be ignored ($l=e$ or $\mu$),
the total decay rate is given by
\begin{equation}
\Gamma\left(B^0\to \pi^- l^+ \nu_l \right) =
\frac{G_F^2|V_{ub}|^2}{192\pi^3m^3_B} \int_0^{q^2_{\rm max}}
dq^2\left[\lambda (q^2)\right]^\frac32 |f^+(q^2)|^2 \label{eq:gamma}
\end{equation}
with $\lambda(q^2)=(m^2_B+m^2_\pi-q^2)^2-4m^2_Bm^2_\pi$.

Results are available for partial branching fractions, over bins in
$q^2$. The tagged analyses from CLEO~\cite{Athar:2003yg},
Belle~\cite{Hokuue:2006nr} and BaBar~\cite{Aubert:2006ry} use three
bins, while BaBar's untagged analysis~\cite{Aubert:2005cd} uses five.
CLEO and BaBar combine results for neutral and charged $B$-meson
decays using isospin symmetry, while Belle quote separate values for
$B^0\to \pi^- l^+ \nu_l$ and $B^+\to\pi^0 l^+\nu_l$. For our analysis,
for the three-bin data, we have combined the Belle charged and neutral
$B$-meson results and subsequently combined these with the CLEO and
BaBar results. Since the systematic errors of the three-bin data are
small compared to the statistical ones, we have ignored correlations
in the systematic errors and combined errors in quadrature. For the
five-bin BaBar data, we assumed that the quoted percentage systematic
errors for the partial branching fractions divided by total branching
fraction are representative for the partial branching fractions alone
and, following BaBar, took them to be fully correlated.

To compute partial branching fractions, we have used $\tau_{B^0}=
1/\Gamma_\mathrm{Tot} = \n(1.527\pm
0.008)e-12n\,\mathrm{s}$~\cite{hfag:2006bi} for the $B^0$ lifetime.

We implement the following fitting procedure. Choose a set of
subtraction points spanning the physical range to use in the Omn\`es
formula of equation~(\ref{eq:omn_th}). Now find the best-fit value
of $\modvub$ and the form factor at the subtraction points to match
both theoretical input form factor values and the experimental partial
branching fraction inputs. The $\chi^2$ function for the fit is thus
(this is very similar to the $\chi^2$ minimisation used
in~\cite{Arnesen:2005ez}):
\begin{eqnarray}
\chi^2 &=& \sum_{i,j=1}^{11}
 \left[f^\mathrm{in}_i-\fom{i}\right]
  C^{-1}_{ij}\left[f^\mathrm{in}_j-\fom{j}\right]\nonumber\\
  & & \mbox{} + 
 \sum_{k,l=1}^8
  \left[B_k^\mathrm{in} -
   B_k^{\mbox{\scriptsize Omn\`es}}(\modvub,f_0,f_1,f_2,f_3)\right]
   C^{-1}_{B\,kl}
   \left[B_l^\mathrm{in} -
   B_l^{\mbox{\scriptsize Omn\`es}}(\modvub,f_0,f_1,f_2,f_3)\right],
   \label{eq:chi2}
\end{eqnarray}
where $f^\mathrm{in}_i$ are input LCSR or lattice QCD values for
$f^+(q^2_i)$ and $B^\mathrm{in}_k$ are input experimental partial
branching fractions. Moreover, $\fom i$ is given by
equation~(\ref{eq:omn_th}) with four subtractions $(q^2_i,f^+(q^2_i))$
at $(0,f_0)$, $(\qsqmax/3,f_1)$, $(2\qsqmax/3,f_2)$ and
$(\qsqmax,f_3)$. The branching fractions $B^{\mbox{\scriptsize
Omn\`es}}$ are calculated using $f^{\mbox{\scriptsize Omn\`es}}$. The
fit parameters are $f_0$, $f_1$, $f_2$, $f_3$ and $\modvub$, where the
latter parameter is used when computing $B^{\mbox{\scriptsize
Omn\`es}}$. We have assumed that the lattice QCD form factor values
have independent statistical uncertainties ($\sigma_i$) and
fully-correlated systematic errors ($\epsilon_i$), leading to an
$11\times11$ covariance matrix with three diagonal blocks: the first
$1\times1$ block is for the LCSR result and the subsequent blocks have
the form $C_{ij}=\sigma_i^2 \delta_{ij} + \epsilon_i\epsilon_j$. The
covariance matrix, $C_B$, for the partial branching fraction inputs is
constructed similarly with three diagonal entries for the three-bin
inputs, together with a block for the five-bin inputs. All the inputs
are listed in tables~\ref{tab:lqcd-inputs} and~\ref{tab:expt-inputs}.
\begin{table}
\begin{center}
\begin{tabular}{@{}l>{$}c<{$}>{$}c<{$}@{}}
\hline
\vrule height2.5ex depth0pt width0pt
 & q^2 & f^\mathrm{in}_i \\
 & \gev^2 \\
\hline
LCSR~\cite{LCSR_04_BZ} & 0 & 0.258\pm0.031 \\
\hline
FNAL~\cite{Okamoto:2005zg} & 15.87 & 0.799\pm0.058 \\
 & 18.58 & 1.128\pm0.086 \\
 & 24.09 & 3.263\pm0.324 \\
\hline
\end{tabular}
\hspace{2em}
\begin{tabular}{@{}l>{$}c<{$}>{$}c<{$}@{}}
\hline
\vrule height2.5ex depth0pt width0pt
 & q^2 & f^\mathrm{in}_i \\
 & \gev^2 \\
\hline
HPQCD~\cite{Gulez:2006dt} & 15.23 & 0.649\pm0.063 \\
 & 16.28 & 0.727\pm0.064 \\
 & 17.34 & 0.815\pm0.065 \\
 & 18.39 & 0.944\pm0.066 \\
 & 19.45 & 1.098\pm0.067 \\
 & 20.51 & 1.248\pm0.097 \\
 & 21.56 & 1.554\pm0.156 \\
\hline
\end{tabular}
\end{center}
\caption{Form factor inputs for the $\chi^2$ function defined in
  equation~(\ref{eq:chi2}). For HPQCD and FNAL the error shown is
  statistical only: the systematic error for input value
  $f^\mathrm{in}_i$ is $ y f^\mathrm{in}_i$, where $y=0.10$ or $0.11$
  respectively. The FNAL inputs are as quoted
  in~\cite{Arnesen:2005ez}.}
\label{tab:lqcd-inputs}
\end{table}
\begin{table}
\begin{center}
\begin{tabular}{@{}l>{$}c<{$}>{$}c<{$}>{$}c<{$}@{}}
\hline
 \vrule height2.5ex depth0pt width0pt
& q^2\mbox{\ range} & 10^4 B^\mathrm{in}_k
 & 10^4 B^{\mbox{\scriptsize Omn\`es}}_k \\
 & \gev^2 \\
\hline
CLEO~\cite{Athar:2003yg}, Belle~\cite{Hokuue:2006nr} 
  &  0$--$8 & 0.410\pm0.056 & 0.451\pm0.041 \\
 \& BaBar~\cite{Aubert:2006ry}
  & 8$--$16 & 0.569\pm0.065 & 0.448\pm0.039 \\
  & >16     & 0.350\pm0.058 & 0.397\pm0.041 \\
\hline
BaBar~\cite{Aubert:2005cd}
  &   0$--$5 & 0.30\pm0.05\pm0.06 & 0.283\pm0.030 \\
  &  5$--$10 & 0.32\pm0.05\pm0.03 & 0.280\pm0.031 \\
  & 10$--$15 & 0.23\pm0.05\pm0.03 & 0.280\pm0.025 \\
  & 15$--$20 & 0.27\pm0.05\pm0.02 & 0.267\pm0.028 \\
  & 20$--$25 & 0.26\pm0.03\pm0.04 & 0.177\pm0.022 \\
\hline
\end{tabular}
\end{center}
\caption{Experimental partial branching fraction inputs for the
  $\chi^2$ function defined in equation~(\ref{eq:chi2}). For the
  partial branching fractions in three bins, the error shown is
  statistical plus systematic combined in quadrature. For the five-bin
  BaBar data, the statistical and systematic errors are shown. We also
  give branching fractions calculated using our fitted form factor and
  $\modvub$.}
\label{tab:expt-inputs}
\end{table}

A fit to the experimental partial branching fractions alone is
sufficient to determine $\modvub f^+(q^2)$. At least one input form
factor value is required in order to extract a result for $\modvub$,
but we have used a set of theoretical inputs to reduce the final error
on the fitted quantities and avoid relying on a single theoretical
calculation.

\section{Results and Discussion}

The best-fit parameters and their Gaussian correlation matrix are:
\begin{equation}
\begin{array}{rcl}
\modvub &=& \vubresult \\
f^+(0) \equiv f_0 &=& \fzeroresult \\
f^+(\qsqmax/3) \equiv f_1 &=& 0.374\pm0.041 \\
f^+(2\qsqmax/3) \equiv f_2 &=& 0.938\pm0.066 \\ 
f^+(\qsqmax) \equiv f_3 &=& 6.63\pm1.28
\end{array}
\qquad
\left(
\begin{array}{ccccc}
1 & -0.31 & -0.86 & -0.77 & -0.52 \\
  &  1    &  0.04 &  0.39 & -0.15 \\
  &       &  1    &  0.67 &  0.65 \\
  &       &       &  1    &  0.24 \\
  &       &       &       &  1
\end{array}
\right)
\end{equation}
The fit has $\chi^2/\mathrm{dof} = 1.1$ for $14$ degrees of freedom.

\begin{figure}
\begin{center}
\includegraphics[width=0.7\hsize]{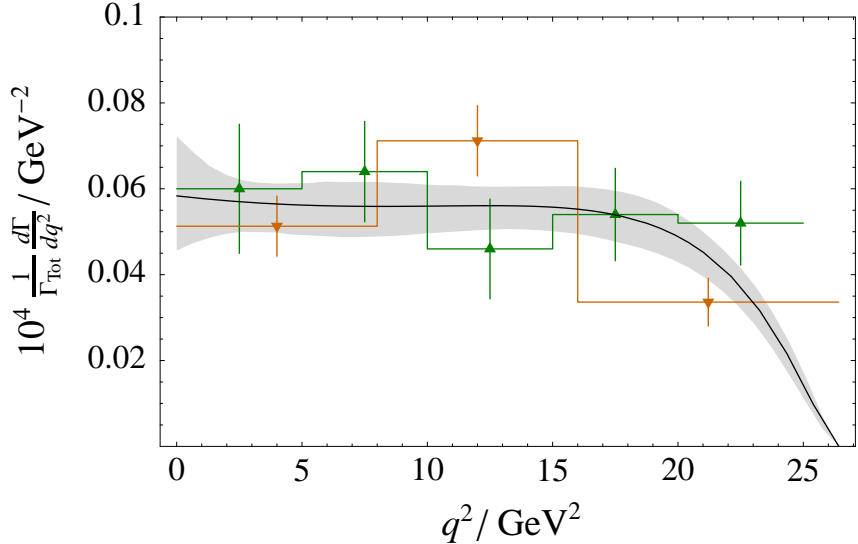}
\end{center}
\caption{Differential decay rate with $68\%$ CL band (shaded) together
  with experimental partial branching fractions divided by the
  appropriate bin-width (histograms and points). Downward triangles
  denote combined CLEO/Belle/BaBar tagged analysis results, upward
  triangles BaBar untagged results.}
\label{fig:ddr}
\end{figure}

In figure~\ref{fig:ddr} we show the differential decay rate calculated
using our fitted form factor and $\modvub$. Partial branching
fractions calculated for the same bins as used experimentally are
given in the last column of table~\ref{tab:expt-inputs}. Our calculated
total branching ratio turns out to be $\n(1.3\pm0.08)e-4n$, in good
agreement with $\n(1.34\pm0.08\pm0.08)e-4n$ quoted by the Heavy
Flavours Averaging Group (HFAG)~\cite{hfag:2006bi}.

\begin{figure}
\begin{center}
\includegraphics[width=0.7\hsize]{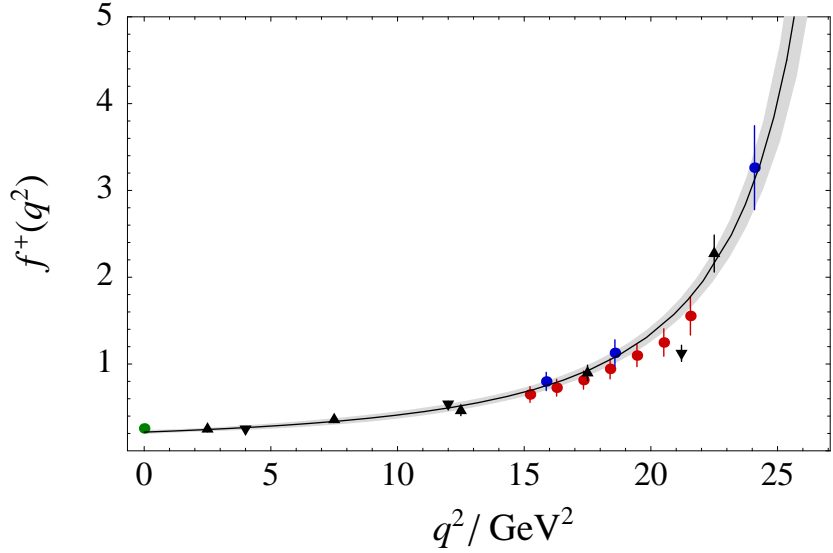}
\end{center}
\caption{Form factor $f^+(q^2)$ with $68\%$ CL band (shaded) together
  with LCSR and lattice QCD inputs (circles). Downward
  (CLEO/Belle/BaBar) and upward (BaBar) triangles show estimates for
  the form factors deduced from the experimental partial branching
  fractions assuming a constant $f^+$ over each bin and using our
  central fitted value of $\modvub$.}
\label{fig:ff}
\end{figure}
\begin{figure}
\begin{center}
\includegraphics[width=0.7\hsize]{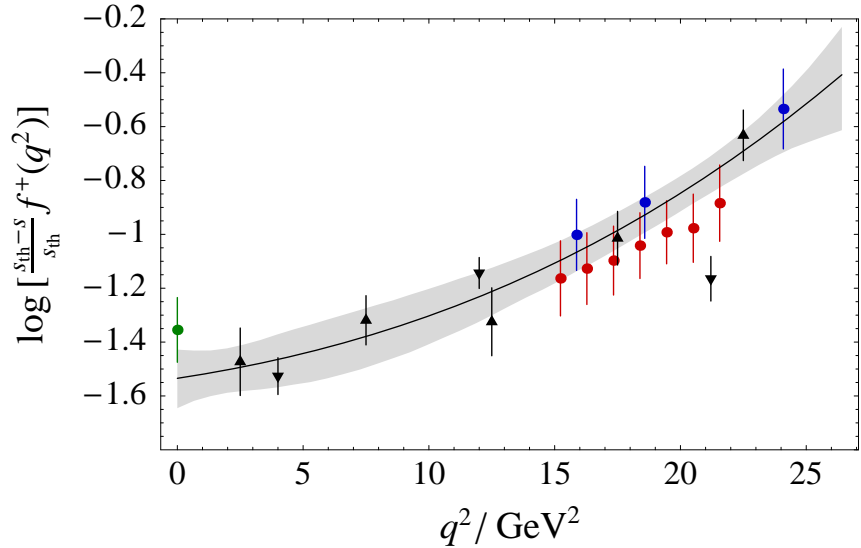}
\end{center}
\caption{Same as in figure~\ref{fig:ff} but for the quantity
$\log[(\sth-s)f^+(q^2)/\sth]$.}
\label{fig:fflog}
\end{figure}
\begin{figure}
\begin{center}
\includegraphics[width=0.7\hsize]{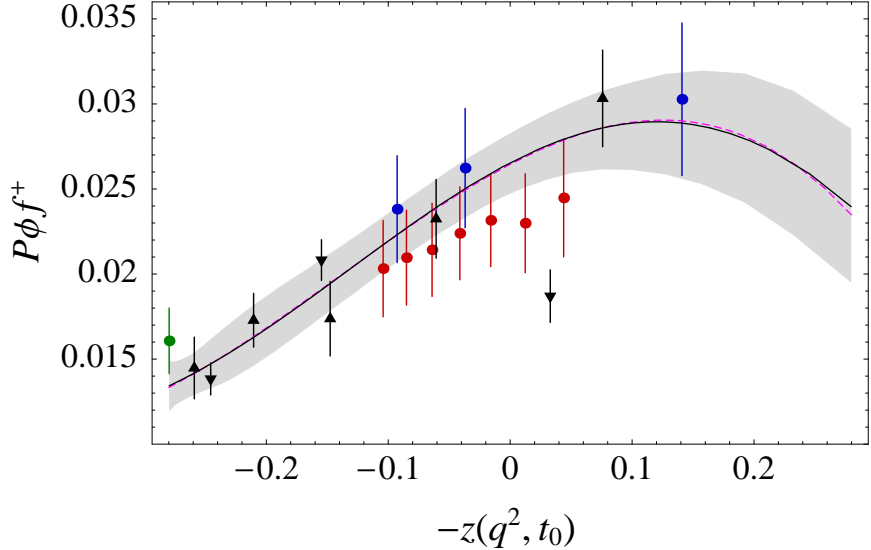}
\end{center}
\caption{Same as in figure~\ref{fig:ff} but for the quantity $P\phi
  f^+$ plotted as a function of $-z(q^2,t_0)$. The dashed line sitting
  on top of the central line is a cubic polynomial fit to $P\phi f^+$,
  see text and equation~(\ref{eq:zfit}).}
\label{fig:ffzz}
\end{figure}

In figure~\ref{fig:ff} we show the form factor $f^+$.
Figure~\ref{fig:fflog} shows the quantity
$\log[(\sth-s)f^+(q^2)/\sth]$ where the details of the fit and inputs
can be better seen. Incorporating the experimental information still
allows a fit which is perfectly consistent with the theory form factor
inputs. Note that the ``experimental'' points (shown by triangles) in
figures~\ref{fig:ff}, \ref{fig:fflog} and \ref{fig:ffzz} are obtained
from the partial branching fractions by assuming a constant form
factor over the corresponding bin and are included as a guide for
convenience. The deviation from our curves of the highest $q^2$-bin
CLEO/Belle/BaBar form factor point is not significant since the form
factor varies rapidly in this region and the calculated partial
branching fraction agrees within errors with the experimental one (as
shown in table~\ref{tab:expt-inputs}).

The inclusion of experimental shape information has balanced the
tendency for the LCSR point at $q^2=0$ to reduce the value of
$\modvub$. To illutrate this, using only the theory inputs and
comparing to the total branching fraction allows the fitted form
factor to pass through the LCSR point and leads to $\modvub =
\n(3.73\pm0.51\pm0.16)e-3n$, where the first error comes from the fit
and the second error is from the HFAG total branching fraction quoted
above. Moreover calculated partial branching fractions from this fit
are above experiment at low $q^2$ and below it at high $q^2$.

We have checked that our determination of $f^+$ is consistent with the
dispersive bound. We computed $P\phi f^+$ as a function of
$z(q^2,t_0)$, where $P$, $\phi$, $z$ and
$t_0=\sth[1-(1-\qsqmax/\sth)^{1/2}]$ are defined in
reference~\cite{Becher:2005bg}\footnote{See equations~(3), (6) and the
intervening text in~\cite{Becher:2005bg}. We use $m_b=4.88\gev$ and
$m_{B^*}=5.235\gev$.}. This is shown in figure~\ref{fig:ffzz}. When
$P\phi f^+$ is Taylor-expanded in powers of $z$, the constraint is
that the sum of squares of the expansion coefficients is bounded above
by $1$. We find that a cubic polynomial is an excellent fit (see
figure~\ref{fig:ffzz}) and the coefficients are,
\begin{equation}
a_0 =  0.026\pm0.002,\quad
a_1 = -0.037\pm0.021,\quad
a_2 = -0.103\pm0.041,\quad
a_3 =  0.25\pm0.37.
\label{eq:zfit}
\end{equation}
with $\sum a_i^2 =0.10^{+0.35}_{-0.06}<1$. The errors for the $a_i$
coefficients arise from the variation of our form factor Monte-Carlo
propagated to $P\phi f^+$ (see the bands in figure~\ref{fig:ffzz}).

One may wonder how important the inclusion of the LCSR point is for
the fit. Removing this input leads to \nolcsr, so $\modvub$ increases
by $6\%$, half its error, while the error itself increases by $15\%$.
Moreover, we checked that the output percentage error in $\modvub$
would decrease about one-eighth as fast as the percentage error on the
LCSR input decreases. Hence the LCSR input is important for its effect
on the central value, but the overall error in $\modvub$ is not much
reduced. The key to the small overall error, as noted
in~\cite{Arnesen:2005ez,Becher:2005bg,Hill:2006ub} is to use a
model-independent functional form with enough parameter freedom to
allow the data to determine the form-factor shape. The Omn\`es form is
relatively simple and is conveniently expressed in terms of form
factor inputs at a set of $q^2$ values.

We have not included possible statistical correlations within and
between the HPQCD and FNAL lattice inputs (the lattice analysis
produces statistical correlations between the form factor values at
different $q^2$, while both simulations are based on the same gauge
field ensembles, although they use different heavy-quark formalisms).
We modelled correlations of the statistical errors both within and
between the HPQCD and FNAL inputs by creating a statistical error
matrix
\[
C_{\mathrm{stat}\,ij} = r \sigma_i \sigma_j +
 (1-r) \sigma_i^2 \delta_{ij}
\]
where $r$ is a correlation coefficient and $\sigma_i$ are the
statistical errors on the individual inputs quoted by the HPQCD and
FNAL groups. We added this to the block-diagonal systematic error
matrix to create the full covariance matrix. For $r=0.25$ our fit
results are essentially unchanged, while for $r=0.81$, the central
value of $\modvub$ moves down by one third of the original error (away
from the inclusive determination) while the error itself grows by
$10\%$. We conclude that these correlations should be included if they
are known, but unless they are strong, they will not have a
substantial effect.

On the experimental side, we have replaced the inputs used here with
partial branching fraction data from BaBar in $12$ bins of
$q^2$~\cite{Aubert:2006fv}, for which full correlation matrices are
given. We find results completely consistent with those given above,
but do not quote them since the data in~\cite{Aubert:2006fv} are still
preliminary.

Applying soft collinear effective theory (SCET) to $B\to\pi\pi$ decays
allows a factorisation result to be derived which leads to a
model-independent extraction of the form factor (multiplied by
$\modvub$) at $q^2=0$~\cite{vubfplus-fact-2004}. We quote the result
from our fit:
\begin{equation}
\modvub f^+(0) = \vubfresult
\end{equation}
to be compared to $\modvub f^+(0) = \n(7.2\pm1.8)e-4n$
in~\cite{vubfplus-fact-2004}. In view of this, we have tried replacing
the LCSR input at $q^2=0$ with the $\modvub f^+(0)$ constraint from
SCET. The result here, \scet, is completely compatible with that using
the lattice inputs alone (\nolcsr). The SCET and LCSR points are not
really compatible with each other when combined separately with the
lattice inputs. Not surprisingly, the effects are larger on $f^+(0)$
than on $\modvub$. Finally, we also tried using both LCSR and SCET
inputs, for which the results (\lcsrscet) are compatible with our
quoted values above.

To conclude, we have presented a theoretically-based procedure to
analyse exclusive $B\to\pi$ semileptonic decays. Starting from very
general principles we propose a simple parameterization for the form
factor $f^+$, equation~(\ref{eq:omn_th}), requiring as input only
knowledge of the form factor at a set of points. We have used this to
combine theoretical and experimental inputs, allowing a robust
determination of $\modvub$ and of the $q^2$ dependence of the form
factor itself. Our error for $\modvub$ is reduced compared to the
current exclusive world-average value,
$\modvub=\n(3.80\pm0.27\pm0.47)e-3n$, from HFAG~\cite{hfag:2006bi} and
is competitive in precision with the inclusive world-average value,
$\modvub=\n(4.45\pm0.20\pm0.26)e-3n$~\cite{hfag:2006bi}. Moreover we
do not find a discrepancy between our exclusive result and the
inclusive world average.

\subsubsection*{Acknowledgements}

We thank Iain Stewart for discussions on dispersive bounds. JMF
acknowledges PPARC grant PP/D000211/1, the hospitality of the
Universidad de Granada and the Institute for Nuclear Theory at the
University of Washington, and thanks the Department of Energy for
partial support. JN acknowledges the hospitality of the School of
Physics \& Astronomy at the University of Southampton, Junta de
Andalucia grant FQM0225, MEC grant FIS2005--00810 and MEC financial
support for movilidad de Profesores de Universidad espa\~noles
PR2006--0403.

\bibliographystyle{physrev}
\bibliography{omnes2}

\end{document}